\documentclass[aps,preprint,showpacs,groupedaddress]{revtex4-1}
\usepackage{amssymb}
\usepackage{mathrsfs}
\usepackage{amsmath}
\usepackage{mathtools}
\usepackage{pstricks}
\usepackage{color}
\usepackage{graphicx}
\usepackage{amsthm}

\usepackage{physics}

\begin{document}


\title{\bf Squeezing as a probe of the universality hypothesis}



\author{P. A. L. Mour\~ao}
\email{pedroaugusto.lima2@outlook.com}
\author{H. A. S. Costa}
\email{hascosta@ufpi.edu.br}
\author{P. R. S. Carvalho}
\email{prscarvalho@ufpi.edu.br}
\affiliation{\it Departamento de F\'\i sica, Universidade Federal do Piau\'\i, 64049-550, Teresina, PI, Brazil}





\begin{abstract}
We compute analytically the radiative quantum corrections, up to next-to-leading loop order, to the universal critical exponents for both massless and massive O($N$) $\lambda\phi^{4}$ scalar squeezed field theories for probing the universality hypothesis. For that, we employ six distinct and independent methods. The outcomes for the universal squeezed critical exponents obtained through these methods are identical among them and reduce to the conventional ones where squeezing is absent. Although the squeezing mechanism modifies the internal properties of the field, the squeezed critical indices are not affected by the squeezing effect, thus implying the validity of the universality hypothesis, at least at the loop level considered. At the end, we present the corresponding physical interpretation for the results in terms of the geometric symmetry properties of the squeezed field.
\end{abstract}


\maketitle


\section{Introduction}

\par Universality shows to be one of the fundamental pillars for the modern of modern phase transitions and critical phenomena. In fact, for obtaining precise values of the critical exponents for a system undergoing a continuous phase transition, we have to employ one of the most powerful mathematical tools of the past century, namely the renormalization group technique \cite{RevModPhys.47.773}. It was designed for treating physical phenomena where all length scales must be considered. This is the situation where the anomalous dimensions emerge. We have to compute anomalous dimensions for many distinct systems. One of the most interesting achievements in the study of phase transitions and critical phenomena is the observation that completely distinct physical systems, as a fluid and a ferromagnet, near a continuous phase transition, display the same critical behavior. This common behavior is then universal and we say that the different systems belong to the same universality class \cite{Stanley}. A given universality class is characterized by a set of universal critical exponents. The critical exponents depend on common properties of the systems as their dimension $d$, $N$ and the symmetry of some $N$-component order parameter if the interactions of their constituents are of short- or long-range type. The universality class studied here is the O($N$) one.  It is a generalization that encompasses the following models for short-range interactions: Ising ($N=1$), XY ($N=2$), Heisenberg ($N=3$), self-avoiding random walk ($N=0$), spherical ($N \rightarrow \infty$) etc. \cite{Pelissetto2002549}. The critical exponents, on the other hand, do not depend on the details of the systems as their critical temperatures $T_{c}$, the form of their lattices etc. The order parameter indicates an ordered state when it assumes nonvanishing values (for $T < T_{c}$) and a disordered one (for $T \geq T_{c}$) when it is zero. The order parameter goes continuously to zero as the critical temperature is approached. One example of an order parameter is magnetization, for magnetic systems. In the field-theoretic momentum-space renormalization group approach employed here \cite{Wilson197475}, the order parameter is identified with the mean value of a fluctuating scalar $N$-component quantum field $\phi$ and its squared mass $m^{2}$ is associated to the difference $T - T_{c}$. Thus a system at the critical temperature $T_{c}$ is described by massless theories. On the other hand for systems at temperatures above $T_{c}$ we have to employ massive theories. The critical exponent values can be computed in an approximation known as the Landau or mean field approximation, where the fluctuations of the quantum field are neglected. In this approximation, the critical exponents results are not precise. We can obtain precise outcomes for the critical exponents only if we take into account the fluctuations of the quantum field. In this work we have to apply the latter approach. For that, we have to evaluate the critical exponents from the scaling properties of the, bare or initially divergent, primitively $1$PI vertex parts, namely the $\Gamma_{B}^{(2)}$, $\Gamma_{B}^{(4)}$ and $\Gamma_{B}^{(2,1)}$ functions. These functions are related to the correlation functions of the theory and we have to compute the critical exponents up to the next-to-leading order (NLO). As the perturbative expansion is divergent, it has to be renormalized, and we renormalize the theory through six distinct and independent methods, by employing three methods for the massless theory and three other methods for the massive one. The advantage of employing six methods is that we can check and compare the results obtained through them. In the field-theoretic formulation employed here, the universality of the critical exponents requires that that their values have to be the same when evaluated through different renormalization methods. In the present work we have to exploit the effect of squeezing in the critical exponents values, instead of the dimension $d$ \cite{PhysRevB.86.155112,PhysRevE.71.046112}, the number of components of the order parameter $N$ \cite{PhysRevLett.110.141601,Butti2005527,PhysRevB.54.7177} and symmetry \cite{PhysRevE.78.061124,Trugenberger2005509}, by defining a field theory for a squeezed fluctuating scalar $N$-component quantum field. Then the universal critical properties of squeezed scalar fields can be studied through their universal anomalous dimensions or, in the phase transitions and critical phenomena language, the universal critical exponents evaluation.
 
\par In Sec. \ref{Squeezed free propagator} we obtain the squeezed free propagator. In both Secs. \ref{Massless theories} and \ref{Massive theories} we compute the universal squeezed critical exponents at and above the critical temperature, respectively. At the end, in Sec. \ref{Conclusions} we display our conclusions and project further works that can result from the present one.

\section{Squeezed free propagator}\label{Squeezed free propagator}

\par We consider a squeezed real scalar quantum field, interacting with itself through the quartic interaction $\lambda\phi^{4}$ where $\lambda = u\kappa^{\epsilon/2}$ is the coupling constant of the theory, $u$ is the dimensionless coupling constant, $\kappa$ is some arbitrary nonvanishing momentum scale value, $\epsilon = 4 - d$ and $d$ is the dimension of the spacetime,
 
\begin{eqnarray}
\phi_{S} = S^{\dagger}(\xi)\phi S(\xi),
\end{eqnarray}
where    
\begin{eqnarray}
S(\xi) = \exp\Bigg[\frac{1}{2}(\xi a^{2} - \xi^{*} a^{\dagger 2})\Bigg]
\end{eqnarray}
is the squeezing operator and $\xi = re^{i\theta}$, where $r$ is the squeezing parameter and $0 \leq r < \infty$ and $0 \leq \theta \leq 2\pi$ \cite{GerryKnight,ScullyZubairy}. Squeezing occurs for $\theta = 0$ or $\theta = \pi$ \cite{GerryKnight} and  we can obtain a squeezed field by choosing any of these values of $\theta$. Without loss of generality, we have to choose the former one. Now we write the quantum field as a single-mode superposition as
\begin{eqnarray}
\phi(\vec{x},t) \equiv \phi(x) = \int \frac{d^{3}k}{(2\pi)^{3}}\frac{1}{\sqrt{2\omega_{k}}}(a_{k}e^{-ikx} + a_{k}^{\dagger}e^{ikx}),
\end{eqnarray}
where $kx = k_{0}t - \vec{k}\cdot\vec{x}$ and $k_{0} = \omega_{k} = \sqrt{\vec{k}^{2} + m^{2}}$. Then we use the fact that, for $\theta = 0$ and thus $\xi = r$, we can easily squeeze a single mode as \cite{Lvovsky1,Lvovsky2,Semmler:16} 
\begin{eqnarray}
S^{\dagger}(r)a_{k}S(r) = a_{k}\cosh (r) - a_{k}^{\dagger}\sinh (r),
\end{eqnarray}
\begin{eqnarray}
S^{\dagger}(r)a_{k}^{\dagger}S(r) = a_{k}^{\dagger}\cosh (r) - a_{k}\sinh (r).
\end{eqnarray}
Now the position-space squeezed free propagator in Minkowski space-time  
\begin{eqnarray}
G_{0,M}(x_{1}, x_{2}) = \langle 0 |T(\phi(x_{1})\phi(x_{2}))| 0 \rangle
\end{eqnarray}
is given by  
\begin{eqnarray}
G_{0,M}(x_{1}, x_{2}) = \int\frac{d^{4}k}{(2\pi)^{4}}G_{0,M}(k)e^{ik(x_{1} - x_{2})}.
\end{eqnarray}
We have discarded, for consistency, the terms that violate causality. Then the $G_{0,M}(k)$ function 
\begin{eqnarray}
G_{0,M}(k) = \frac{i\cosh (2r)}{k^{2} - m^{2} + i\varepsilon}
\end{eqnarray}
is the momentum-space squeezed free propagator in Minkowski space-time. As it is known, for field theory computations in critical phenomena, it is convenient to employ the computations by using the free propagator in its Euclidean version $G_{0,E}(k)$. It is obtained from that in  Minkowski space-time through a Wick rotation. Thus omitting the subscript $E$ we have 
\begin{eqnarray}
G_{0,E}(k) \equiv G_{0}(k) = \frac{\cosh (2r)}{k^{2} + m^{2}}.
\end{eqnarray}
Now we compute the effect of squeezing on the critical exponents through the application of six distinct and independent methods.

\section{Massless theories}\label{Massless theories}

\par At the critical temperature the three bare primitively $1$PI vertex parts, up to NLO, to be renormalized are
\begin{eqnarray}
\Gamma^{(2)}_{B} = \frac{k^{2}}{\cosh (2r)} - \frac{1}{6}\hspace{1mm}\parbox{12mm}{\includegraphics[scale=1.0]{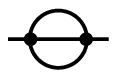}} - \frac{1}{4}\hspace{1mm}\parbox{10mm}{\includegraphics[scale=0.9]{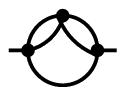}} ,
\end{eqnarray}
\begin{eqnarray}
&& \Gamma^{(4)}_{B} = - \hspace{1mm}\parbox{10mm}{\includegraphics[scale=0.09]{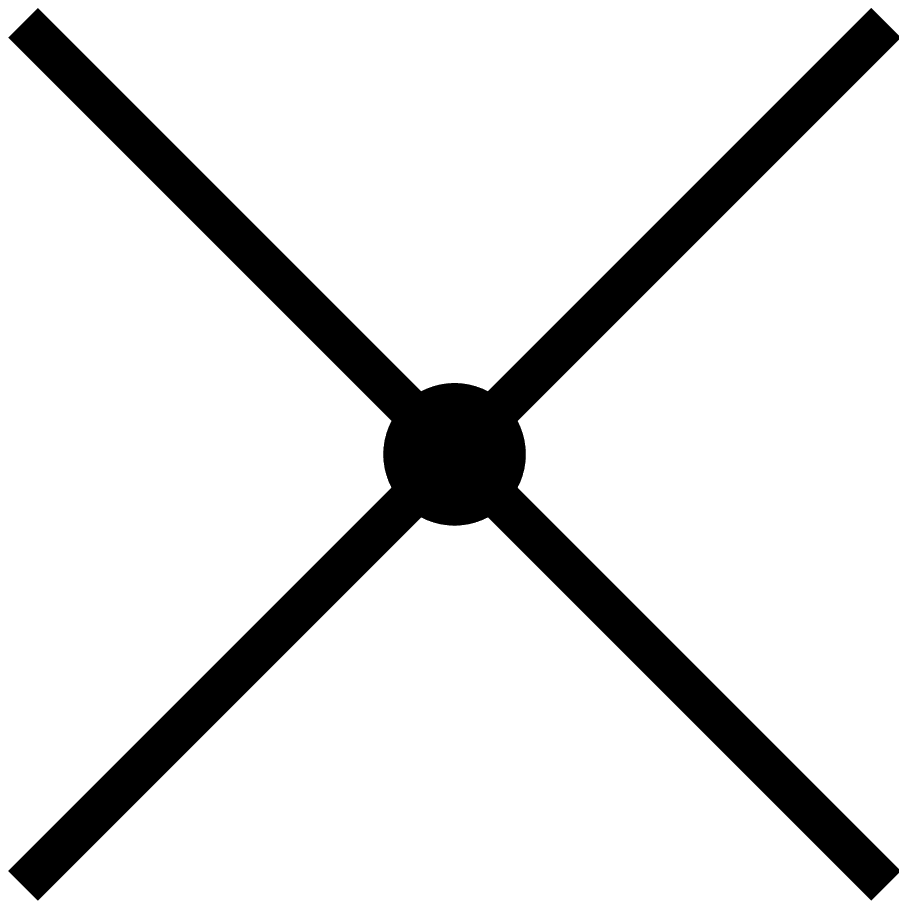}} - \frac{1}{2}\hspace{1mm}\Biggr(\parbox{10mm}{\includegraphics[scale=1.0]{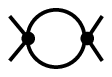}} + 2 \hspace{1mm} perm.\Biggr)  - \frac{1}{4}\hspace{1mm}\Biggr(\parbox{16mm}{\includegraphics[scale=1.0]{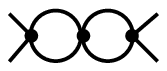}} + 2 \hspace{1mm} perm.\Biggr)  -  \nonumber \\&& \frac{1}{2}\hspace{1mm}\Biggr(\parbox{12mm}{\includegraphics[scale=0.8]{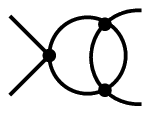}} + 5 \hspace{1mm} perm.\Biggr),
\end{eqnarray}
\begin{eqnarray}
\Gamma^{(2,1)}_{B} = \parbox{14mm}{\includegraphics[scale=1.0]{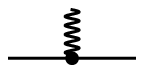}} - \frac{1}{2}\hspace{1mm}\parbox{14mm}{\includegraphics[scale=1.0]{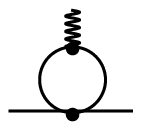}} - \frac{1}{4}\hspace{1mm}\parbox{12mm}{\includegraphics[scale=1.0]{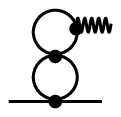}} - \frac{1}{2}\hspace{1mm}\parbox{12mm}{\includegraphics[scale=0.8]{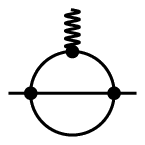}},
\end{eqnarray}
where \textit{perm.} means a permutation of the external momenta attached to the cut external lines. The internal lines $\parbox{12mm}{\includegraphics[scale=1.0]{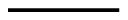}} \equiv G_{0}(k)\vert_{m^{2} = 0}$ represent the massless squeezed free propagator $G_{0}(k)\vert_{m^{2} = 0} = \cosh (2r)/k^{2}$, the wave line in Eq. (12) is associated to composite fields and $\Gamma_{B}^{(2,1)}$ is the bare $1$PI vertex part corresponding to composite fields. We consider only $\Gamma_{B}^{(2,1)}$ and not higher ones. The former are primitive $1$PI vertex parts for composite fields and the higher ones can be written in terms of the latter. This means that once $\Gamma_{B}^{(2,1)}$ is renormalized, the higher ones get automatically renormalized as well. We apply the multiplicative renormalization to the bare primitively $1$PI vertex parts through the following renormalization constants 
\begin{eqnarray}
\Gamma_{R}^{(n, l)}(P_{i}, Q_{j}, u, \kappa) = Z_{\phi_{S}}^{n/2}Z_{\phi_{S}^{2}}^{l}\Gamma_{B}^{(n, l)}(P_{i}, Q_{j}, \lambda_{B}),
\end{eqnarray}
$(i = 1, \cdots, n, j = 1, \cdots, l)$. The renormalized $1$PI vertex parts satisfy to the renormalization group equation
\begin{eqnarray}
\left( \kappa\frac{\partial}{\partial\kappa} + \beta_{S}\frac{\partial}{\partial u} - \frac{1}{2}n\gamma_{\phi_{S}} + l\gamma_{\phi_{S}^{2}} \right)\Gamma_{R}^{(n, l)} = 0
\end{eqnarray}
where $\kappa$ is some arbitrary momentum scale parameter, $\epsilon = 4 - d$,
\begin{eqnarray}
\beta_{S}(u) = \kappa\frac{\partial u}{\partial \kappa} = -\epsilon\left(\frac{\partial\ln u_{0}}{\partial u}\right)^{-1},
\end{eqnarray}
\begin{eqnarray}
\gamma_{\phi_{S}}(u) = \beta_{S}(u)\frac{\partial\ln Z_{\phi_{S}}}{\partial u},
\end{eqnarray}
\begin{eqnarray}
\gamma_{\phi_{S}^{2}}(u) = -\beta_{S}(u)\frac{\partial\ln Z_{\phi_{S}^{2}}}{\partial u},
\end{eqnarray}
and we have to employ
\begin{eqnarray}
\overline{\gamma}_{\phi_{S}^{2}}(u) = -\beta_{S}(u)\frac{\partial\ln \overline{Z}_{\phi_{S}^{2}}}{\partial u} \equiv \gamma_{\phi_{S}^{2}}(u) - \gamma_{\phi_{S}}(u)
\end{eqnarray}
in the next two methods of this Section, for convenience, where $\overline{Z}_{\phi_{S}^{2}} \equiv Z_{\phi_{S}}Z_{\phi_{S}^{2}}$. In the next two renormalization methods applied we start from the bare primitively $1$PI vertex parts while in the last method we start from the renormalized ones.

\subsection{Normalization conditions method}

\par In the normalization conditions method \cite{Amit,BrezinLeGuillouZinnJustin}, the external momenta of the needed Feynman diagrams are held at fixed nonvanishing values
\begin{eqnarray}
\parbox{10mm}{\includegraphics[scale=1.0]{fig10.eps}}_{SP} \equiv \parbox{10mm}{\includegraphics[scale=1.0]{fig10.eps}}\Bigg\vert_{P^{2} = \kappa^{2}},
\end{eqnarray}
\begin{eqnarray}
\parbox{12mm}{\includegraphics[scale=1.0]{fig6.eps}}^{\prime} \equiv \frac{\partial}{\partial P^{2}}\parbox{11mm}{\includegraphics[scale=1.0]{fig6.eps}} \Bigg\vert_{P^{2} = \kappa^{2}},
\end{eqnarray}
\begin{eqnarray}
\parbox{12mm}{\includegraphics[scale=.8]{fig21.eps}}_{SP} \equiv \parbox{12mm}{\includegraphics[scale=0.8]{fig21.eps}}\Bigg\vert_{P^{2} = \kappa^{2}},
\end{eqnarray}
\begin{eqnarray}
\parbox{10mm}{\includegraphics[scale=0.9]{fig7.eps}}^{\prime} \equiv \frac{\partial}{\partial P^{2}}\parbox{12mm}{\includegraphics[scale=0.9]{fig7.eps}} \Bigg\vert_{P^{2} = \kappa^{2}},
\end{eqnarray}
where SP is the symmetry point $P_{i}\cdot P_{j} = (\kappa^{2}/4)(4\delta_{ij}-1)$, implying that $(P_{i} + P_{j})^{2} \equiv P^{2} = \kappa^{2}$ for $i\neq j$. Now by evaluating the squeezed $\beta_{S}$-function and the squeezed field $\gamma_{\phi ,S}$ and squeezed composite field $\overline{\gamma}_{\phi^{2},S}$ anomalous dimensions for arbitrary values of $N$, with $\Gamma_{B}^{(2)} \rightarrow q\Gamma_{B}^{(2)}$ and $\Gamma^{(2)} \rightarrow q\Gamma^{(2)}$, we get 
\begin{eqnarray}\label{a}
\beta_{S}(u) = -\epsilon u +   \frac{N + 8}{6}\left( 1 + \frac{1}{2}\epsilon \right)\cosh^{2}(2r) u^{2} -  \frac{3N + 14}{12}\cosh^{4}(2r)u^{3}, 
\end{eqnarray}
\begin{eqnarray}\label{b}
\gamma_{\phi ,S}(u) = \frac{N + 2}{72}\left( 1 + \frac{5}{4}\epsilon \right)\cosh^{4}(2r)u^{2} -  \frac{(N + 2)(N + 8)}{864}\cosh^{6}(2r)u^{3},  
\end{eqnarray}
\begin{eqnarray}\label{c}
\overline{\gamma}_{\phi^{2}, S}(u) = \frac{N + 2}{6}\left( 1 + \frac{1}{2}\epsilon \right)\cosh^{2}(2r) u -  \frac{N + 2}{12}\cosh^{4}(2r)u^{2},
\end{eqnarray}
where $u$ is the dimensionless renormalized coupling constant. We see that the $\beta_{S}$-function and anomalous dimensions just above assume finite values, as it is demanded by any renormalization scheme, and depend on the symmetry point value employed as it is seen in the second, first, and first terms of Eqs. (\ref{a}), (\ref{b}) and (\ref{c}), respectively. Since the symmetry point is not a universal feature, the squeezed critical exponents will not depend on this parameter as we will show. The nontrivial root or nontrivial fixed point ($u^{*} \neq 0$) of the $\beta_{S}$-function is needed for the computation of the radiative quantum corrections to the critical exponents, while the trivial one ($u^{*} = 0$) is used in the critical indices evaluation in the mean field or Landau approximation. Thus the nontrivial fixed point is given by  
\begin{eqnarray}
u^{*} = \frac{6\epsilon}{(N + 8)\cosh^{2}(2r)}  \Biggl \{ 1 + \epsilon\Biggl[ \frac{3(3N + 14)}{(N + 8)^{2}} - \frac{1}{2}  \Biggr] \Biggr \}.
\end{eqnarray}
Once we have obtained the nontrivial fixed point, we can compute two of the squeezed critical exponents through the following relations $\eta_{S}\equiv\gamma_{\phi ,S}(u^{*})$ and $\nu_{S}^{-1}\equiv 2 - \eta_{S} - \overline{\gamma}_{\phi^{2}, S}(u^{*})$. The remaining four squeezed critical indices can be obtained through four scaling relations \cite{Stanley}. 

\subsection{Minimal subtraction scheme}

\par In this renormalization scheme, we start from the bare theory as in the earlier method. On the other hand, now the external momenta of the Feynman diagrams are kept at arbitrary values \cite{Amit,BrezinLeGuillouZinnJustin} then making this renormalization method an elegant and general one. We have not applied the more popular $\overline{MS}$-scheme, since nonuniversal extra factors (ex. $\ln 4\pi$, Euler's constant $\gamma$ etc.) can be absorbed in the definitions of the coupling constant and do not produce any physical effect. The squeezed $\beta_{S}$-function and the squeezed field $\gamma_{\phi ,S}$ and squeezed composite field $\overline{\gamma}_{\phi^{2},S}$ anomalous dimensions assume the form
\begin{eqnarray}\label{beta}
\beta_{S}(u) = -\epsilon u +   \frac{N + 8}{6}\cosh^{2}(2r) u^{2} - \frac{3N + 14}{12}\cosh^{4}(2r)u^{3}, 
\end{eqnarray}
\begin{eqnarray}\label{gamma}
\gamma_{\phi ,S}(u) = \frac{N + 2}{72}\cosh^{4}(2r)u^{2} - \frac{(N + 2)(N + 8)}{864}\cosh^{6}(2r)u^{3},  
\end{eqnarray}
\begin{eqnarray}
\overline{\gamma}_{\phi^{2}, S}(u) = \frac{N + 2}{6}\cosh^{2}(2r) u - \frac{N + 2}{12}\cosh^{4}(2r)u^{2}.
\end{eqnarray}
The squeezed $\beta_{S}$-function and anomalous dimensions do not depend on external momenta values since they have canceled out in the middle of calculations. The corresponding nontrivial fixed point is given by 
\begin{eqnarray}
u^{*} = \frac{6\epsilon}{(N + 8)\cosh^{2}(2r)} \Biggl\{ 1 + \epsilon\Biggl[ \frac{3(3N + 14)}{(N + 8)^{2}} \Biggr]\Biggr\}.
\end{eqnarray}
Now by computing the two squeezed critical exponents $\eta_{S}\equiv\gamma_{\phi ,S}(u^{*})$ and $\nu_{S}^{-1}\equiv 2 - \eta_{S} - \overline{\gamma}_{\phi^{2}, S}(u^{*})$, we obtain the same results as obtained through the earlier renormalization method thus confirming that the referred critical exponents are universal quantities. Distinct renormalization methods have produced the same results and this serves as a check of the results through independent methods. Now we proceed to calculate the squeezed critical exponents through the last method employed in this Section.

\subsection{Massless BPHZ method}

\par As opposed to the earlier methods, in the present renormalization scheme, namely the BPHZ (Bogoliubov-Parasyuk-Hepp-Zimmermann) \cite{BogoliubovParasyuk,Hepp,Zimmermann,Kleinert}, we start from the renormalized theory, instead from the bare one. The operator $\mathcal{K}$ extracts only the divergent term of a given diagram and $\mathcal{R}$ subtracts both logarithmically and quadratically divergent subdiagrams \cite{Kleinert}. The theory is renormalized by the following renormalization constants 

\begin{eqnarray}
Z_{\phi_{S}} = 1 + \frac{1}{P^2} \Biggl[ \frac{1}{6} \mathcal{K}\mathcal{R} 
\left(\parbox{12mm}{\includegraphics[scale=1.0]{fig6.eps}}
\right) + \frac{1}{4} \mathcal{K}\mathcal{R} 
\left(\parbox{10mm}{\includegraphics[scale=0.9]{fig7.eps}} \right)  + \frac{1}{3} \mathcal{K}\mathcal{R}
  \left(\parbox{12mm}{\includegraphics[scale=1.0]{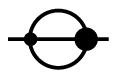}} \right) \Biggr],
\end{eqnarray}

\begin{eqnarray}
&& Z_{u} = 1 + \frac{1}{\kappa^{\epsilon}u} \Biggl[ \frac{1}{2} \mathcal{K}\mathcal{R} 
\left(\parbox{10mm}{\includegraphics[scale=1.0]{fig10.eps}} + 2 \hspace{1mm} perm.\right)
 +  \frac{1}{4} \mathcal{K}\mathcal{R} 
\left(\parbox{17mm}{\includegraphics[scale=1.0]{fig11.eps}} + 2 \hspace{1mm} perm.\right)   +  \nonumber \\ && \frac{1}{2} \mathcal{K}\mathcal{R} 
\left(\parbox{12mm}{\includegraphics[scale=.8]{fig21.eps}} + 5 \hspace{1mm} perm.\right)  +  \mathcal{K}\mathcal{R}
  \left(\parbox{10mm}{\includegraphics[scale=1.0]{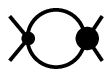}} + 2 \hspace{1mm} perm.\right)  \Biggr],
\end{eqnarray}

\begin{eqnarray}
&& Z_{\phi_{S}^{2}} = 1 + \frac{1}{2} \mathcal{K}\mathcal{R} 
\left(\parbox{14mm}{\includegraphics[scale=1.0]{fig14.eps}} \right)  + \frac{1}{4} \mathcal{K}\mathcal{R} 
\left(\parbox{12mm}{\includegraphics[scale=1.0]{fig16.eps}} \right)  +  \frac{1}{2} \mathcal{K}\mathcal{R} 
\left(\parbox{11mm}{\includegraphics[scale=.8]{fig17.eps}} \right)  + \nonumber \\ && \frac{1}{2} \mathcal{K}\mathcal{R}
  \left(\parbox{12mm}{\includegraphics[scale=.2]{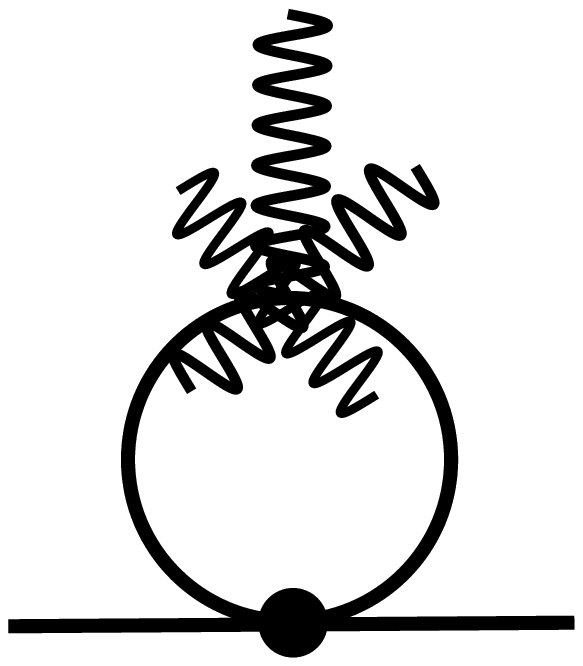}} \right)  +  \frac{1}{2} \mathcal{K}\mathcal{R}
  \left(\parbox{12mm}{\includegraphics[scale=.2]{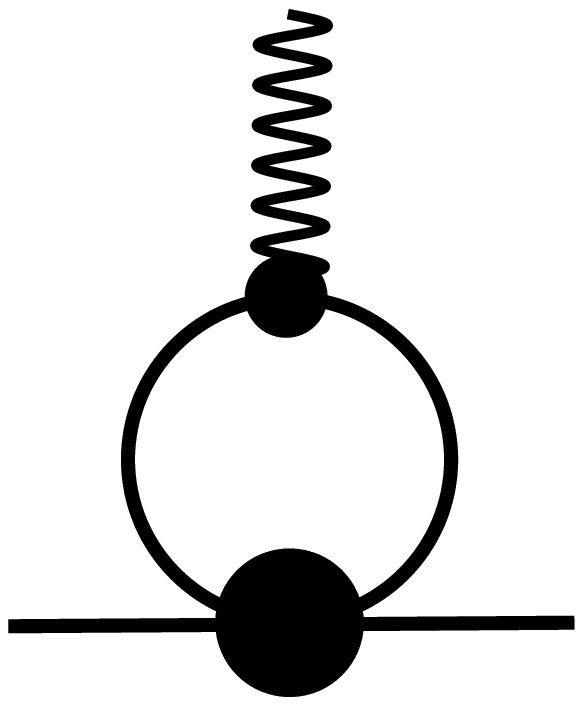}} \right).
\end{eqnarray}

Now we can evaluate the the squeezed $\beta_{S}$-function (the same as Eq. (\ref{beta})) and the squeezed field $\gamma_{\phi ,S}$ (the same as Eq. (\ref{gamma})) and squeezed composite field $\gamma_{\phi^{2},S}$ anomalous dimensions. Thus we obtain
\begin{eqnarray}
\gamma_{\phi^{2}, S}(u) = \frac{N + 2}{6}\cosh^{2}(2r) u - \frac{5(N + 2)}{72}\cosh^{4}(2r)u^{2}.
\end{eqnarray}
Once again, the squeezed $\beta_{S}$-function and anomalous dimensions do not depend on the arbitrary external momenta values. By computing the corresponding nontrivial fixed point we get the same value as that of the earlier method and by applying the relations $\eta\equiv\gamma_{\phi}(u^{*})$ and $\nu^{-1}\equiv 2 - \gamma_{\phi^{2}}(u^{*})$. This shows once again that the critical indices are universal quantities since they have identical values when obtained through three distinct and independent renormalization methods.

\section{Massive theories}\label{Massive theories}

\par Above the critical temperature, we have to consider a massive theory and thus a massive squeezed propagator $\parbox{12mm}{\includegraphics[scale=1.0]{fig9.eps}} \equiv G_{0}(k)$. Now the bare primitively $1$PI vertex parts 
\begin{eqnarray}
&& \Gamma^{(2)}_{B} = \frac{k^{2} + m^{2}}{\cosh (2r)} - \frac{1}{2}\hspace{1mm}\parbox{10mm}{\includegraphics[scale=0.9]{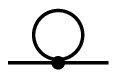}} - \frac{1}{4}\hspace{1mm}\parbox{10mm}{\includegraphics[scale=0.9]{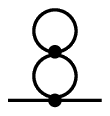}} - \frac{1}{6}\hspace{1mm}\parbox{12mm}{\includegraphics[scale=1.0]{fig6.eps}} - \frac{1}{4}\hspace{1mm}\parbox{12mm}{\includegraphics[scale=0.9]{fig7.eps}} - \nonumber \\ && \frac{1}{4}\hspace{1mm}\parbox{12mm}{\includegraphics[scale=0.9]{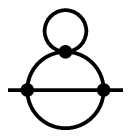}} -  \frac{1}{12}\hspace{1mm}\parbox{10mm}{\includegraphics[scale=0.9]{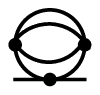}} - \frac{1}{8}\hspace{1mm}\parbox{10mm}{\includegraphics[scale=0.9]{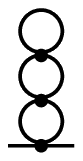}} -  \frac{1}{8}\hspace{1mm}\parbox{12mm}{\includegraphics[scale=0.9]{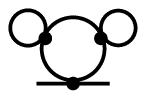}}\quad ,
\end{eqnarray}
\begin{eqnarray}
&& \Gamma^{(4)}_{B} = - \hspace{1mm}\parbox{10mm}{\includegraphics[scale=0.09]{fig29.eps}} - \frac{1}{2}\hspace{1mm}\Biggr(\parbox{10mm}{\includegraphics[scale=1.0]{fig10.eps}} + 2 \hspace{1mm} perm.\Biggr)  -  \frac{1}{4}\hspace{1mm}\Biggr(\parbox{16mm}{\includegraphics[scale=1.0]{fig11.eps}} + 2 \hspace{1mm} perm.\Biggr)  - \nonumber \\ && \frac{1}{2}\hspace{1mm}\Biggr(\parbox{12mm}{\includegraphics[scale=0.8]{fig21.eps}} + 5 \hspace{1mm} perm.\Biggr) -  \frac{1}{2}\hspace{1mm}\Biggr(\parbox{12mm}{\includegraphics[scale=1.0]{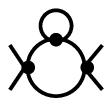}} + 2 \hspace{1mm} perm.\Biggr), 
\end{eqnarray}
\begin{eqnarray}
\Gamma^{(2,1)}_{B} = \parbox{14mm}{\includegraphics[scale=1.0]{fig28.eps}} - \frac{1}{2}\hspace{1mm}\parbox{14mm}{\includegraphics[scale=1.0]{fig14.eps}} - \frac{1}{4}\hspace{1mm}\parbox{12mm}{\includegraphics[scale=1.0]{fig16.eps}} - \frac{1}{2}\hspace{1mm}\parbox{12mm}{\includegraphics[scale=0.8]{fig17.eps}} - \frac{1}{2}\hspace{1mm}\parbox{14mm}{\includegraphics[scale=0.8]{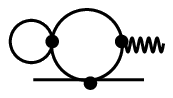}},
\end{eqnarray}
satisfy to the Callan-Symanzik equation
\begin{eqnarray}
&& \left( m\frac{\partial}{\partial m} + \beta_{S}\frac{\partial}{\partial u} - \frac{1}{2}n\gamma_{\phi_{S}} + l\gamma_{\phi_{S}^{2}} \right)\Gamma_{R}^{(n, l)}(P_{i}, Q_{j}, u, m) \nonumber\\ && = m^{2}(2 - \gamma_{\phi_{S}})\Gamma_{R}^{(n, l + 1)}(P_{i}, Q_{j}, 0, u, m)
\end{eqnarray}
where now $m$ is used as the momentum scale parameter and
\begin{eqnarray}
\beta_{S}(u) = m\frac{\partial u}{\partial m} = -\epsilon\left(\frac{\partial\ln u_{0}}{\partial u}\right)^{-1},
\end{eqnarray}
\begin{eqnarray}
\gamma_{\phi_{S}}(u) = \beta_{S}(u)\frac{\partial\ln Z_{\phi_{S}}}{\partial u},
\end{eqnarray}
\begin{eqnarray}
\gamma_{\phi_{S}^{2}}(u) = -\beta_{S}(u)\frac{\partial\ln Z_{\phi_{S}^{2}}}{\partial u},
\end{eqnarray}
where 
\begin{eqnarray}
\overline{\gamma}_{\phi_{S}^{2}}(u) = -\beta_{S}(u)\frac{\partial\ln \overline{Z}_{\phi_{S}^{2}}}{\partial u} \equiv \gamma_{\phi_{S}^{2}}(u) - \gamma_{\phi_{S}}(u).
\end{eqnarray}

\subsection{Callan-Symanzik method}

\par In the this method \cite{Amit,BrezinLeGuillouZinnJustin}, we fix the external momenta of the needed diagrams at vanishing values
\begin{eqnarray}
\parbox{10mm}{\includegraphics[scale=1.0]{fig10.eps}}_{SP} \equiv \parbox{10mm}{\includegraphics[scale=1.0]{fig10.eps}}\Bigg\vert_{P^{2} = 0},
\end{eqnarray}
\begin{eqnarray}
\parbox{12mm}{\includegraphics[scale=1.0]{fig6.eps}}^{\prime} \equiv \frac{\partial}{\partial P^{2}}\parbox{11mm}{\includegraphics[scale=1.0]{fig6.eps}} \Bigg\vert_{P^{2} = 0},
\end{eqnarray}
\begin{eqnarray}
\parbox{12mm}{\includegraphics[scale=.8]{fig21.eps}}_{SP} \equiv \parbox{12mm}{\includegraphics[scale=0.8]{fig21.eps}}\Bigg\vert_{P^{2} = 0},
\end{eqnarray}
\begin{eqnarray}
\parbox{10mm}{\includegraphics[scale=0.9]{fig7.eps}}^{\prime} \equiv \frac{\partial}{\partial P^{2}}\parbox{12mm}{\includegraphics[scale=0.9]{fig7.eps}} \Bigg\vert_{P^{2} = 0}.
\end{eqnarray}
Now we compute the squeezed $\beta_{S}$-function, squeezed field and squeezed composite field anomalous dimensions and obtain
\begin{eqnarray}
\beta_{S}(u) = -\epsilon u +   \frac{N + 8}{6}\left( 1 - \frac{1}{2}\epsilon \right)\cosh^{2}(2r) u^{2} -  \frac{3N + 14}{12}\cosh^{4}(2r)u^{3}, 
\end{eqnarray}
\begin{eqnarray}
&& \gamma_{\phi ,S}(u) = \frac{N + 2}{72}\left( 1 - \frac{1}{4}\epsilon + I\epsilon \right)\cosh^{4}(2r)u^{2} - \nonumber \\ &&   \frac{(N + 2)(N + 8)}{864}( 1 + I )\cosh^{6}(2r)u^{3},  
\end{eqnarray}
\begin{eqnarray}
\overline{\gamma}_{\phi^{2}, S}(u) = \frac{N + 2}{6}\left( 1 - \frac{1}{2}\epsilon \right)\cosh^{2}(2r) u -   \frac{N + 2}{12}\cosh^{4}(2r)u^{2},
\end{eqnarray}
where the integral $I$ is a number given by
\begin{eqnarray}\label{d}
&& I = \int_{0}^{1} \left\{ \frac{1}{1 - x(1 - x)} + \frac{x(1 - x)}{[1 - x(1 - x)]^{2}}\right\}.
\end{eqnarray}
and is just a number \cite{Amit} which cancels out in the renormalization process, where $I \approx 1.4728$. Now by applying the relations $\eta_{S}\equiv\gamma_{\phi ,S}(u^{*})$ and $\nu_{S}^{-1}\equiv 2 - \eta_{S} - \overline{\gamma}_{\phi^{2}, S}(u^{*})$, we obtain that the critical exponents found are the same as that computed through the earlier methods, since the integral $I$ cancels out in the middle of calculations. Now we proceed to apply the next renormalization group method.

\subsection{Unconventional minimal subtraction scheme}

\par In the present method, the external momenta of the Feynman integrals are kept at their arbitrary values \cite{J.Math.Phys.542013093301} and the tadpoles can be eliminated \cite{Amit}. Then we obtain the squeezed $\beta_{S}$-function, squeezed field and squeezed composite field anomalous dimensions. Their expressions are the same as the corresponding ones for the minimal subtraction method. Then by applying the relations $\eta_{S}\equiv\gamma_{\phi ,S}(u^{*})$ and $\nu_{S}^{-1}\equiv 2 - \eta_{S} - \overline{\gamma}_{\phi^{2}, S}(u^{*})$, we obtain that the squeezed critical exponents are identical to those obtained through the earlier methods. 

\subsection{Massive BPHZ method}

\par In this method, we start from the renormalized theory by applying the BPHZ (Bogoliubov-Parasyuk-Hepp-Zimmermann) \cite{BogoliubovParasyuk,Hepp,Zimmermann,Kleinert} renormalization procedure for obtaining the renormalization constants
\begin{eqnarray}
Z_{\phi_{S}} = 1 + \frac{1}{P^2} \Biggl[ \frac{1}{6} \mathcal{K}\mathcal{R} \left(\parbox{12mm}{\includegraphics[scale=1.0]{fig6.eps}} \right) + \frac{1}{4} \mathcal{K}\mathcal{R} \left(\parbox{10mm}{\includegraphics[scale=0.9]{fig7.eps}} \right)  +  \frac{1}{3} \mathcal{K}\mathcal{R} \left(\parbox{12mm}{\includegraphics[scale=1.0]{fig26.eps}} \right) \Biggr],
\end{eqnarray}
\begin{eqnarray}
&& Z_{u} = 1 + \frac{1}{m^{\epsilon}u} \Biggl[ \frac{1}{2} \mathcal{K}\mathcal{R} 
\left(\parbox{10mm}{\includegraphics[scale=1.0]{fig10.eps}} + 2 \hspace{1mm} perm.\right)
 + \frac{1}{4} \mathcal{K}\mathcal{R} \left(\parbox{17mm}{\includegraphics[scale=1.0]{fig11.eps}} + 2 \hspace{1mm} perm.\right)   + \nonumber \\ &&  \frac{1}{2} \mathcal{K}\mathcal{R} \left(\parbox{12mm}{\includegraphics[scale=.8]{fig21.eps}} + 5 \hspace{1mm} perm.\right)  + \frac{1}{2} \mathcal{K}\mathcal{R} \left(\parbox{10mm}{\includegraphics[scale=1.0]{fig13.eps}} + 2 \hspace{1mm} perm.\right) + \nonumber \\ && \mathcal{K}\mathcal{R}
  \left(\parbox{10mm}{\includegraphics[scale=1.0]{fig25.eps}} + 2 \hspace{1mm} perm.\right)  +  \mathcal{K}\mathcal{R}\left(\parbox{10mm}{\includegraphics[scale=1.0]{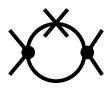}} + 2 \hspace{1mm} perm.\right) \Biggr],
\end{eqnarray}
\begin{eqnarray}\label{Zphi}
&& Z_{m^{2}} = 1 + \frac{1}{m^{2}} \Biggl[ \frac{1}{2} \mathcal{K}\mathcal{R} 
\left(\parbox{11mm}{\includegraphics[scale=1.0]{fig1.eps}}
\right) + \frac{1}{4}\mathcal{K}\mathcal{R}\left(\parbox{10mm}{\includegraphics[scale=1.0]{fig2.eps}} \right)  +  \frac{1}{2}\mathcal{K}\mathcal{R}\left(\parbox{7mm}{\includegraphics[scale=1.0]{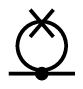}} \right)  +  \nonumber \\ && \frac{1}{2} \mathcal{K}\mathcal{R}\left(\parbox{11mm}{\includegraphics[scale=1.0]{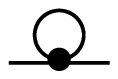}} \right)  +  \frac{1}{6}\mathcal{K}\mathcal{R}\left(\parbox{11mm}{\includegraphics[scale=1.0]{fig6.eps}} \right)\Biggr|_{P^{2} = 0}  \Biggr].
\end{eqnarray}
The squeezed $\beta_{S}$-function and corresponding field anomalous dimensions needed in this method present the same expressions as that for the massless BPHZ method. The mass anomalous dimension is given by
\begin{eqnarray}
\gamma_{m^{2}}(u) = \frac{N + 2}{6}\cosh^{2}(2r) u - \frac{5(N + 2)}{72}\cosh^{4}(2r)u^{2}.
\end{eqnarray}
Then by applying the relations $\eta_{S}\equiv\gamma_{\phi_{S}}(u^{*})$ and $\nu_{S}^{-1}\equiv 2 - \gamma_{m^{2}}(u^{*})$, we obtain the same squeezed critical exponents as obtained by the earlier methods.

\section{Results}

By applying all the aforementioned methods, and equations like $\eta_{S}\equiv\gamma_{\phi_{S}}(u^{*})$ and $\nu_{S}^{-1}\equiv 2 - \gamma_{m^{2}}(u^{*})$, we can compute the squeezed critical exponents values, namely 
\begin{eqnarray}
\alpha_{S} \equiv \alpha = \frac{(4 - N)}{4(N + 8)}\epsilon +  \frac{(N + 2)(N^{2} + 30N + 56)}{4(N + 8)^{3}}\epsilon^{2},
\end{eqnarray}
\begin{eqnarray}
\beta_{S} \equiv \beta = \frac{1}{2} - \frac{3}{2(N + 8)}\epsilon +  \frac{(N + 2)(2N + 1)}{2(N + 8)^{3}}\epsilon^{2},
\end{eqnarray}
\begin{eqnarray}
\gamma_{S} \equiv \gamma = 1 + \frac{(N + 2)}{2(N + 8)}\epsilon +  \frac{(N + 2)(N^{2} + 22N + 52)}{4(N + 8)^{3}}\epsilon^{2},
\end{eqnarray}
\begin{eqnarray}
\delta_{S} \equiv \delta = 3 + \epsilon +  \frac{N^{2} + 14N + 60}{2(N + 8)^{2}}\epsilon^{2},
\end{eqnarray}
\begin{eqnarray}
\nu_{S} \equiv \nu = \frac{1}{2} + \frac{(N + 2)}{4(N + 8)}\epsilon +  \frac{(N + 2)(N^{2} + 23N + 60)}{8(N + 8)^{3}}\epsilon^{2},
\end{eqnarray}
\begin{eqnarray}
\eta_{S} \equiv \eta = \frac{(N + 2)}{2(N + 8)^{2}q}\epsilon^{2}\left\{ 1 + \left[ \frac{6(3N + 14)}{(N + 8)^{2}} -\frac{1}{4} - \frac{2(N + 2)}{(N + 8)^{2}}\frac{(1 - q)}{q} \right]\epsilon\right\}.
\end{eqnarray}
We have found that they are the same as that without the squeezing effect present \cite{Wilson197475}, thus confirming the universality hypothesis, at least at the loop order considered. This result keeps the universality hypothesis intact. It means that although we modify the internal properties of the field, by squeezing it, this mechanism is not capable of modifying its internal symmetries, namely the discrete $\phi \rightarrow -\phi$ and rotational O($N$) ones, on which the universal critical exponents depend, among other parameters as $d$ and $N$ and the range on the interactions. 

\section{Conclusions}\label{Conclusions}

\par In this work we have computed analytically the NLO loop corrections to the critical exponents for both massless and massive O($N$) $\lambda\phi^{4}$ scalar squeezed field theories. For that, we have employed six distinct and independent renormalization group schemes. We have shown that the outcomes for the squeezed critical exponents are the same as those computed through the six methods, thus showing that they are universal quantities. This shows the great utility of employing more that one method since we can compare the results obtained by applying distinct methods. Squeezing does not affect the critical exponents. The corresponding physical interpretation is that although the internal properties of the field are modified by the squeezing mechanism, this mechanism is not capable of modifying the internal symmetries of the field, namely the discrete $\phi \rightarrow -\phi$ and rotational O($N$) ones, on which the universal critical exponents depend, among other parameters as $d$ and $N$ and the range of the interactions. Moreover the present work enables us to probe the effect of squeezing in the outcomes for critical exponents for finite systems, bulk amplitude ratios as well as corrections to scaling etc.

\section{Acknowledgements}

\par PALM, HASC and PRSC would like to thank CAPES and CNPq (Brazilian funding agencies) for financial support. Particularly, PRSC would like to thank CNPq for grants: Universal-$431727/2018$ and Produtividade 307982/2019-0.

\section{Appendix}

\subsection{Results for each Feynman graph needed in the Normalization conditions method}

\begin{eqnarray}
\parbox{10mm}{\includegraphics[scale=1.0]{fig10.eps}}_{SP} = \frac{1}{\epsilon}\left(1 + \frac{1}{2}\epsilon  \right)\cosh^{2}(2r), 
\end{eqnarray}   
\begin{eqnarray}
\parbox{12mm}{\includegraphics[scale=1.0]{fig6.eps}}^{\prime} = -\frac{1}{8\epsilon}\left( 1 + \frac{5}{4}\epsilon \right)\cosh^{3}(2r),
\end{eqnarray}  
\begin{eqnarray}
\parbox{10mm}{\includegraphics[scale=0.9]{fig7.eps}}^{\prime} = -\frac{1}{6\epsilon^{2}}\left( 1 + 2\epsilon  \right)\cosh^{5}(2r),
\end{eqnarray}  
\begin{eqnarray}
\parbox{12mm}{\includegraphics[scale=0.8]{fig21.eps}}_{SP} = \frac{1}{2\epsilon^{2}}\left( 1 + \frac{3}{2}\epsilon \right)\cosh^{4}(2r).
\end{eqnarray} 

\subsection{Results for each Feynman graph needed in the Minimal subtraction scheme}

\begin{eqnarray}
\parbox{10mm}{\includegraphics[scale=1.0]{fig10.eps}} = \frac{1}{\epsilon} \left[1 - \frac{1}{2}\epsilon - \frac{1}{2}\epsilon L(P^{2}) \right]\cosh^{2}(2r),
\end{eqnarray}   
\begin{eqnarray}
\parbox{12mm}{\includegraphics[scale=1.0]{fig6.eps}} =  -\frac{P^{2}}{8\epsilon}\left[ 1 + \frac{1}{4}\epsilon -2\epsilon L_{3}(P^{2}) \right]\cosh^{3}(2r),
\end{eqnarray}  
\begin{eqnarray}
\parbox{10mm}{\includegraphics[scale=0.9]{fig7.eps}} =  -\frac{P^{2}}{6\epsilon^{2}}\left[ 1 + \frac{1}{2}\epsilon -3\epsilon L_{3}(P^{2}) \right]\cosh^{5}(2r),
\end{eqnarray}  
\begin{eqnarray}
\parbox{12mm}{\includegraphics[scale=0.8]{fig21.eps}} = \frac{1}{2\epsilon^{2}}\left[1 - \frac{1}{2}\epsilon - \epsilon L(P^{2}) \right]\cosh^{4}(2r),
\end{eqnarray}
where $P$ are the external momenta and
\begin{eqnarray}\label{uhduhguh}
L(P^{2}) = \int_{0}^{1}dx\ln[x(1-x)P^{2}],
\end{eqnarray}
\begin{eqnarray}\label{uhduhguhf}
L_{3}(P^{2}) = \int_{0}^{1}dx(1-x)\ln[x(1-x)P^{2}].
\end{eqnarray}

\subsection{Results for each Feynman graph needed in the Massless BPHZ method}

As it is known, in the massless theory, we need just a minimal set of Feynman diagrams \cite{Amit}. They are 
\begin{eqnarray}
\parbox{10mm}{\includegraphics[scale=1.0]{fig10.eps}} =  \frac{\mu^{\epsilon}u^{2}}{\epsilon} \left[1 - \frac{1}{2}\epsilon - \frac{1}{2}\epsilon L\Bigg(\frac{P^{2}}{\mu^{2}}\Bigg) \right]\cosh^{2}(2r),
\end{eqnarray}
\begin{eqnarray}
\parbox{12mm}{\includegraphics[scale=1.0]{fig6.eps}} =  -\frac{P^{2}u^{2}}{8\epsilon}\left[ 1 + \frac{1}{4}\epsilon -2\epsilon L_{3}\Bigg(\frac{P^{2}}{\mu^{2}}\Bigg) \right]\cosh^{3}(2r),
\end{eqnarray}  
\begin{eqnarray}
\parbox{10mm}{\includegraphics[scale=0.9]{fig7.eps}} = \frac{P^{2}u^{3}}{6\epsilon^{2}}\left[ 1 + \frac{1}{2}\epsilon -3\epsilon L_{3}\Bigg(\frac{P^{2}}{\mu^{2}}\Bigg) \right]\cosh^{5}(2r),
\end{eqnarray}      
\begin{eqnarray}
\parbox{12mm}{\includegraphics[scale=0.8]{fig21.eps}} =  -\frac{\mu^{\epsilon}u^{3}}{2\epsilon^{2}} \left[1 - \frac{1}{2}\epsilon - \epsilon L\Bigg(\frac{P^{2}}{\mu^{2}}\Bigg) \right]\cosh^{4}(2r).
\end{eqnarray}  

\subsection{Results for each Feynman graph needed in the Callan-Symanzik method}

\begin{eqnarray}
&&\parbox{10mm}{\includegraphics[scale=1.0]{fig10.eps}}_{SP} = \frac{1}{\epsilon}\left(1 - \frac{1}{2}\epsilon \right)\cosh^{2}(2r),
\end{eqnarray}   
\begin{eqnarray}
&&\parbox{12mm}{\includegraphics[scale=1.0]{fig6.eps}}^{\prime} = -\frac{1}{8\epsilon}\left( 1 - \frac{1}{4}\epsilon +I\epsilon \right)\cosh^{3}(2r),
\end{eqnarray}  
\begin{eqnarray}
&&\parbox{12mm}{\includegraphics[scale=0.9]{fig7.eps}}^{\prime} = -\frac{1}{6\epsilon^{2}}\left( 1 - \frac{1}{4}\epsilon +\frac{3}{2}I\epsilon \right)\cosh^{5}(2r),
\end{eqnarray}  
\begin{eqnarray}
&&\parbox{12mm}{\includegraphics[scale=0.8]{fig21.eps}}_{SP} = \frac{1}{2\epsilon^{2}}\left(1 - \frac{1}{2}\epsilon \right)\cosh^{4}(2r),
\end{eqnarray}  
where the integral $I$ is given in Eq. (\ref{d}).

\subsection{Results for each Feynman graph needed in the Unconventional minimal subtraction scheme}

\begin{eqnarray}
\parbox{10mm}{\includegraphics[scale=1.0]{fig10.eps}} = \frac{1}{\epsilon} \left[1 - \frac{1}{2}\epsilon - \frac{1}{2}\epsilon L(P^{2}, m_{B}^{2}) \right]\cosh^{2}(2r),
\end{eqnarray}   

\begin{eqnarray}
&& \parbox{12mm}{\includegraphics[scale=1.0]{fig6.eps}} = \left\{-\frac{3 m_{B}^{2}}{2 \epsilon^{2}}\left[1 + \frac{1}{2}\epsilon + \left(\frac{\pi^{2}}{12} +1 \right)\epsilon^{2} \right] - \frac{3 m_{B}^{2}}{4}\tilde{i}(P^{2}, m_{B}^{2})  - \right.  \nonumber \\&&  \left. \frac{P^{2}}{8 \epsilon}\left[1 + \frac{1}{4}\epsilon - 2 \epsilon L_{3}(P^{2},m_{B}^{2})\right]\right\}\cosh^{3}(2r), 
\end{eqnarray}
\begin{eqnarray}
&& \parbox{12mm}{\includegraphics[scale=1.0]{fig7.eps}} = \left\{-\frac{5 m_{B}^{2}}{3 \epsilon^{3}}\left[1 + \epsilon + \left(\frac{\pi^{2}}{24} + \frac{15}{4} \right)\epsilon^{2} \right] - \frac{5 m_{B}^{2}}{2 \epsilon}\tilde{i}(P^{2}, m_{B}^{2}) - \right.  \nonumber \\&&  \left. \frac{P^{2}}{6 \epsilon^{2}}\left[1+ \frac{1}{2}\epsilon - 3 \epsilon L_{3}(P^{2},m_{B}^{2})\right]\right\}\cosh^{5}(2r), 
\end{eqnarray}
\begin{eqnarray}
\parbox{14mm}{\includegraphics[scale=1.0]{fig21.eps}} = \frac{1}{\epsilon^{2}} \left[1 - \frac{1}{2}\epsilon - \epsilon L(P^{2}, m_{B}^{2}) \right]\cosh^{4}(2r),
\end{eqnarray}  
where
\begin{eqnarray}
L(P^{2}, m_{B}^{2}) = \int_{0}^{1}dx\ln[x(1-x)P^{2} + m_{B}^{2}],
\end{eqnarray}
\begin{eqnarray}
L_{3}(P^{2}, m_{B}^{2}) = \int_{0}^{1}dx(1-x)\ln[x(1-x)P^{2} + m_{B}^{2}],
\end{eqnarray}
\begin{eqnarray}
&& \tilde{i}(P^{2}, m_{B}^{2}) = \int_{0}^{1} dx \times \nonumber \\&& \int_{0}^{1}dy\ln y \frac{d}{dy}\left((1-y)\ln\left\{y(1-y)P^{2} + \left[1-y + \frac{y}{x(1-x)}\right]m_{B}^{2} \right\}\right),
\end{eqnarray}

\subsection{Results for each Feynman graph needed in the Massive BPHZ method}

\begin{eqnarray}
\left(\parbox{12mm}{\includegraphics[scale=1.0]{fig6.eps}}\right)\Biggr|_{m^{2}=0} =  -\frac{P^{2}u^{2}}{8\epsilon}\left[ 1 + \frac{1}{4}\epsilon -2\epsilon\, J_{3}(P^{2}) \right]\cosh^{3}(2r),
\end{eqnarray}
\begin{eqnarray}
\parbox{12mm}{\includegraphics[scale=1.0]{fig7.eps}}\bigg|_{m^{2}=0} = \frac{P^{2}u^{3}}{6\epsilon^{2}}\left[1 + \frac{1}{2}\epsilon - 3\epsilon\, J_{3}(P^{2})\right]\cosh^{5}(2r),
\end{eqnarray}
\begin{eqnarray}
\parbox{10mm}{\includegraphics[scale=1.0]{fig26.eps}} \quad = -\frac{3P^{2}u^{3}}{16\epsilon^{2}}\left[1 + \frac{1}{4}\epsilon - 2\epsilon\, J_{3}(P^{2})\right]\cosh^{5}(2r),
\end{eqnarray}
\begin{eqnarray}
\parbox{10mm}{\includegraphics[scale=1.0]{fig10.eps}} = \frac{\mu^{\epsilon}u^{2}}{\epsilon} \left[1 - \frac{1}{2}\epsilon - \frac{1}{2}\epsilon J(P^{2}) \right]\cosh^{2}(2r),
\end{eqnarray}
\begin{eqnarray}
\parbox{16mm}{\includegraphics[scale=1.0]{fig11.eps}} = -\frac{\mu^{\epsilon}u^{3}}{\epsilon^{2}} \left[1 - \epsilon - \epsilon J(P^{2}) \right]\cosh^{4}(2r),
\end{eqnarray}
\begin{eqnarray}
\parbox{12mm}{\includegraphics[scale=0.8]{fig21.eps}} = -\frac{\mu^{\epsilon}u^{3}}{2\epsilon^{2}} \left[1 - \frac{1}{2}\epsilon - \epsilon J(P^{2}) \right]\cosh^{4}(2r),
\end{eqnarray}
\begin{eqnarray}
\parbox{12mm}{\includegraphics[scale=1.0]{fig13.eps}} =   \frac{\mu^{\epsilon}u^{3}}{2\epsilon^{2}} J_{4}(P^{2})\cosh^{4}(2r),
\end{eqnarray}
\begin{eqnarray}
\parbox{10mm}{\includegraphics[scale=1.0]{fig25.eps}} = \frac{3\mu^{\epsilon}u^{3}}{2\epsilon^{2}} \left[1 - \frac{1}{2}\epsilon - \frac{1}{2}\epsilon J(P^{2}) \right]\cosh^{4}(2r),
\end{eqnarray}
\begin{eqnarray}
\parbox{12mm}{\includegraphics[scale=1.0]{fig24.eps}} =  -\frac{\mu^{\epsilon}u^{3}}{2\epsilon^{2}} J_{4}(P^{2})\cosh^{4}(2r),
\end{eqnarray}
\begin{eqnarray}
\parbox{12mm}{\includegraphics[scale=1.0]{fig1.eps}} =
\frac{m^{2}u}{(4\pi)^{2}\epsilon}\left[ 1 - \frac{1}{2}\epsilon\ln\left(\frac{m^{2}}{\mu^{2}}\right)\right]\cosh(2r),
\end{eqnarray}
\begin{eqnarray}
\parbox{8mm}{\includegraphics[scale=1.0]{fig2.eps}} = - \frac{m^{2}u^{2}}{\epsilon^{2}}\left[ 1 - \frac{1}{2}\epsilon - \epsilon\ln\left(\frac{m^{2}}{\mu^{2}}\right)\right]\cosh^{3}(2r),
\end{eqnarray}
\begin{eqnarray}
\parbox{12mm}{\includegraphics[scale=1.0]{fig22.eps}} =  \frac{m^{2}u^{2}}{2\epsilon^{2}}\left[ 1 - \frac{1}{2}\epsilon - \frac{1}{2} \epsilon\ln\left(\frac{m^{2}}{\mu^{2}}\right)\right]\cosh^{3}(2r),
\end{eqnarray}
\begin{eqnarray}
\parbox{12mm}{\includegraphics[scale=1.0]{fig23.eps}} =  \frac{3m^{2}u^{2}}{2\epsilon^{2}}\left[ 1 - \frac{1}{2} \epsilon\ln\left(\frac{m^{2}}{\mu^{2}}\right)\right]\cosh^{3}(2r),
\end{eqnarray}
\begin{eqnarray}
\left(\parbox{12mm}{\includegraphics[scale=1.0]{fig6.eps}}\right)\Biggr|_{P^{2}=0} = -\frac{3m^{2}u^{2}}{2\epsilon}\left[ 1 + \frac{1}{2}\epsilon -\epsilon\ln\left(\frac{m^{2}}{\mu^{2}}\right)\right]\cosh^{3}(2r),
\end{eqnarray}
where
\begin{eqnarray}\label{uhduhufgjg}
J(P^{2}) = \int_{0}^{1}dx \ln \left[\frac{x(1-x)(P^{2}) + m^{2}}{\mu^{2}}\right],
\end{eqnarray}
\begin{eqnarray}\label{uhduhufgjgdhg}
J_{3}(P^{2}) = \int_{0}^{1}\int_{0}^{1}dxdy\,(1-y)\ln \Biggl\{\frac{y(1-y)(P^{2})}{\mu^{2}} + \left[1-y + \frac{y}{x(1-x)}  \right]\frac{m^{2}}{\mu^{2}}\Biggr\}, \nonumber \\  
\end{eqnarray}
\begin{eqnarray}\label{ugujjgdhg}
J_{4}(P^{2}) = \frac{m^{2}}{\mu^{2}}\int_{0}^{1}dx\frac{(1 - x)}{\frac{x(1 - x)(P^{2})}{\mu^{2}} + \frac{m^{2}}{\mu^{2}}}.
\end{eqnarray}

\bibliography{apstemplate}

\end{document}